\documentclass[useAMS,usenatbib]{mn2e}
\usepackage{amssymb}
\usepackage{graphicx}

\def\kms{{\rm \thinspace km \thinspace s}^{-1}}

\def\Lsun{\hbox{$\rm\thinspace L_{\odot}$}}

\def\Msun{\hbox{$\rm\thinspace M_{\odot}$}}
\def\pc{{\rm\thinspace pc}}     
       
\def\yr{{\rm\thinspace yr}}
\def\kyr{{\rm\thinspace kyr}}
\def\Myr{{\rm\thinspace Myr}}

\def\K{{\rm\thinspace K}}

\def\ndens{\thinspace\mathrm{cm}^{-3}}

\def\flux{{\rm \thinspace\gamma\thinspace s}^{-1}}
\def\fitnorm{{\rm \thinspace cm^{-3} \thinspace s\thinspace \gamma}^{-1}}

\title{The Return of the Proplyds - Understanding the Dynamics of Ionization Triggered Stars}

\author[Matthias
Gritschneder \& Andreas Burkert]{Matthias Gritschneder$^{1}$\thanks{E-mail:
    gritschn@ucolick.org},Andreas Burkert$^{2,3}$\\
$^1$Astronomy and Astrophysics Department, University of California, Santa Cruz,
  CA 95064, USA \\
$^2$ University Observatory Munich, Scheinerstrasse 1, 81679 Munich,
Germany \\
$^3$ Max-Planck-Fellow, Max-Planck-Institute for Extraterrestrial
Physics, Giessenbachstrasse 1, 85758 Garching, Germany} 

\begin{document}

\maketitle

\label{firstpage}

\begin{abstract}
Proplyds and stars inside HII-regions are a well studied
phenomenon. It is possible that they were triggered by
the expansion of the HII-region itself.
Here, we present calculations on the dynamics of HII-regions. We
show that the triggered stars that form in the expanding shell of
swept up material around the HII region rarely return into the HII
regions on timescales that are inferred for the proplyds and observed
young stars. However, in very dense environments like Orion, the
triggered stars return in time. Thus,  our model can explain why
proplyds are barely observed in other HII regions.
We propose that the properties of young stellar objects in HII regions
in general depend critically on the distance from the massive,
ionizing central star cluster. 
Closest in, there are proplyds, where the
disk of a young star interacts directly with the feedback of the
massive star. Further out are Class II protostars, where the
ionization already removed the envelope. Even further away, one should
find Class I stars, which either have been triggered by the ionizing
radiation or pre-existed and have not lost their envelope yet. This
radial sequence is not necessarily an age sequence but rather a result
of the dwindling importance of stellar winds and ionizing radiation with
distance.
We investigate the observational signature of triggered star formation
and find that the stellar distribution for ionization triggered star
formation shows a distinct
feature, a peak at the current position of the ionization front.
Therefore, it is generally possible to tell triggered
and in situ distributions of stars apart.\end{abstract} 


\vspace{-.1cm}
\section{Introduction}
HII-regions and their constituents have been at the fore-front of
astronomic research for a long time
\citep[e.g.][]{Stromgren:1939lr,Elmegreen:1977fx,Spitzer:1978fk}. As
they are bright, they lend themselves easily to observations. Within them,
a host of different objects are observed, ranging from giant pillars
\citep{Hester:1996fk} and peculiar gaseous structures to faint, evaporating globules. In
the era of the Spitzer Space Telescope, a vast number of young stellar objects (YSOs) are
discovered in and around them. Generally, the morphology and idealized
evolution of HII regions are well understood
\citep[e.g.][]{Stromgren:1939lr,Spitzer:1978fk}. 
The location of YSOs in HII regions provide important
information on their origin. However, up to now, little attention has
been given to the question of how the YSOs have reached their current
location. Here, we attempt an analytic approach to investigate the
heritage of the YSOs. 

Among the YSOs, the so-called proplyds
(\citealt{Laques:1979uq,Odell:1993uq}, short for 
protoplanetary disk) are particularly interesting. These are young
protostellar disks, illuminated by an O-star 
in direct proximity. They are mainly found in Orion. Due to their
relative proximity, their ionized envelopes and their disks can be
observed in the optical with the Hubble Space Telescope
\citep[e.g.][]{Odell:1993uq,Bally:2005vn,De-Marco:2006}. All of those
objects close to the Trapezium Cluster show near-IR counterparts
\citep{Meaburn:1988uq,Smith:2005fk}, indicative of star
formation.
\citet{Odell:1998zr} finds a weak correlation of
proplyd size with the distance to the ionizing source in Orion. 

Still, they are barely found in other HII-regions than
Orion. This can be
explained if they are extremely transient objects.
\citet{Smith:2005fk} show that the proplyds are very concentrated
around the youngest ionization sources and have ages less than
$500\kyr$. They already concluded that further out, only the remnants
of proplyds, e.g. stars with very light disks are observed. Here, we
test this scenario further with dynamical considerations.

Recent studies find a host of proplyd like features in various
HII-regions (e.g. \citealt{Smith:2003qf} in the Carina Nebula and \citealt{Wright:2012kx} in
Cygnus, further away from the ionizing source). However, nearly all of them 
have no IR-counterparts. It remains to be seen if they are failed
proplyds or are at an earlier stage, leading to proplyd
formation. Estimations of the photo-evaporation efficiency support the
former assumption \citep[e.g.][]{Henney:1999kx}.

Detailed simulations of the formation and evolution of proplyds exist
\citep[e.g.][]{Storzer:1999ys,Garcia-Arredondo:2001bh,Henney:2009}. However,
a complete picture of their dynamics is still
missing. An interesting question is whether they formed from stars
triggered at the edge of the HII region that later on returned into
the region, either by falling behind the expanding shell or by
turning around due to gravity.

Here, we focus on the dynamical aspect, the rather straightforward
question how a system of YSOs that were triggered to form in the
expanding shell surrounding an HII region evolves and where these
young stars during and after their proplyd phase would be found observationally. 
We review the basic equations in \S 2 and present a calculation of
representative trajectories in \S 3. In \S 4 we investigate the
question how big a observed sample has to be to allow for
distinguishing different origins and in \S 5 we draw the conclusions.
\vspace{-.1cm}
\section{Basic Equations}
\label{equations}
\begin{figure*}
\begin{center}
{\centering 
\includegraphics[width=15cm]{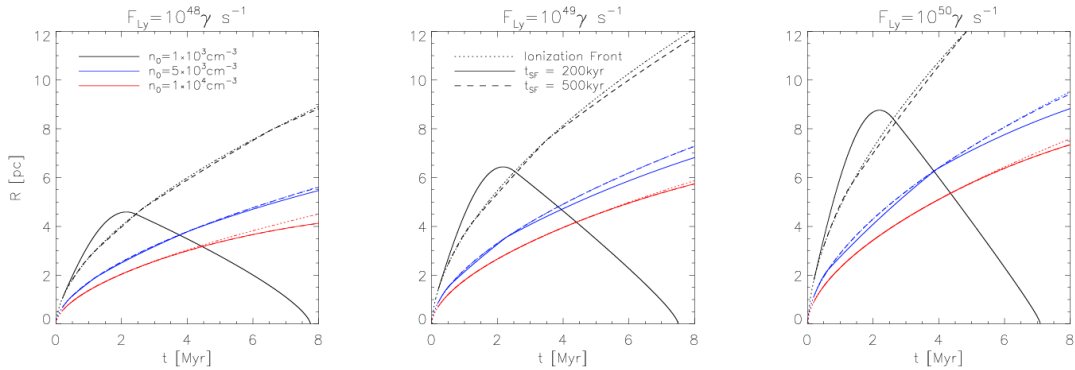}}
\end{center}
\caption{Trajectories of stars formed inside the ionization shock
  front at different times with different initial densities and
  fluxes. Dotted black lines show the evolution of the ionisation
  front. Solid and dashed lines correspond to the orbits of stars that
  formed after $200\kyr$ and $500\kyr$, respectively. The stars move under
  the influence of the gravitational force of the gas and a central
  cluster of $M_{\rm central}=100\Msun$. 
    \label{fig:100Msun}}
\end{figure*}
\begin{figure*}
\begin{center}
{\centering 
\includegraphics[width=15cm]{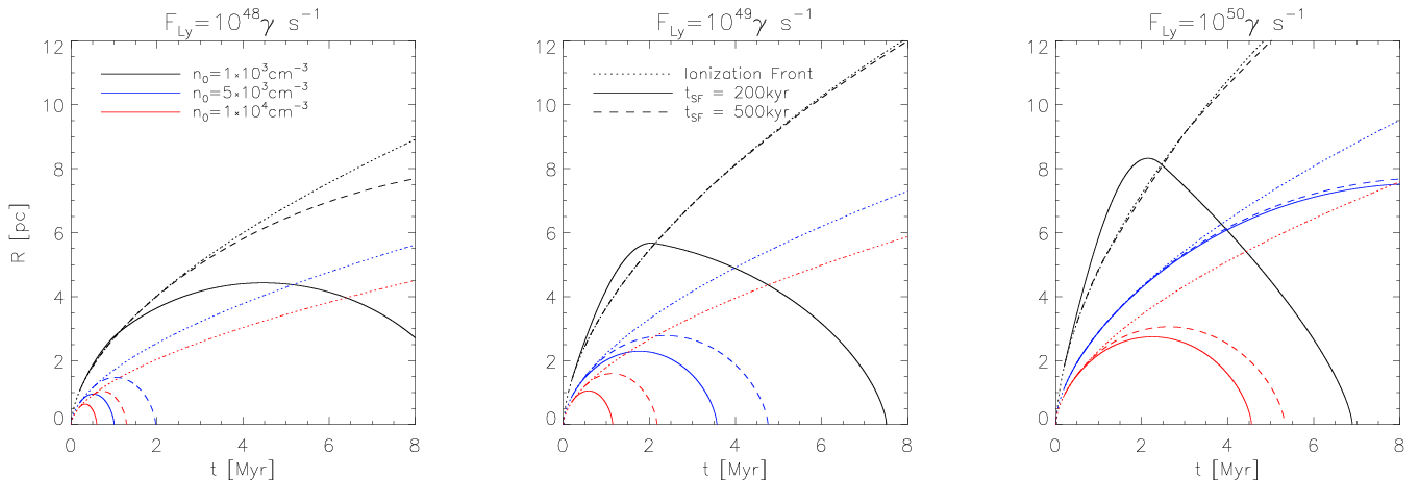}}
\end{center}
\caption{Same as Figure \ref{fig:100Msun}, but for $M_{\rm central}=1000\Msun$.
    \label{fig:1000Msun}}
\end{figure*}

When a massive star ionizes its surrounding, the photoionization leads
to a heating of the gas. The radius of a sphere which can be
immediately ionized by a source with a flux of Lyman photons $J_{\rm Ly}$ is given by the
Str{\"o}mgren radius
\begin{equation}
R_{\rm S} = (\frac{3 J_{\rm Ly}}{4\pi n_0^2\alpha_{\rm B}})^{(1/3)},
\end{equation}
where $n_0$ is the number density of the ionized gas and $\alpha_{\rm
  B}$ is the recombination coefficient.
Scattering between the photoelectrons and the ions and cooling
processes results in an equilibrium temperature $T_{\rm ion}$ inside
the sphere. Due to the increased pressure inside, the HII region
expands into the cold surrounding. In a medium of
constant density the radius at a given time t is given as
\begin{equation}
\label{R_t}
R_{\rm f}(t)=R_\mathrm{s}\left(1+\frac{7}{4}\frac{c_\mathrm{s,hot}}{R_\mathrm{s}}(t-t_0)\right)^\frac{4}{7},
\end{equation}
where $c_\mathrm{s,hot}$ is the sound speed of the hot, ionized
gas. The velocity of
the shell at a time t is
\begin{equation}
\label{v_t}
v_{\rm f}(t)=c_\mathrm{s,hot}\left(1+\frac{7}{4}\frac{c_\mathrm{s,hot}}{R_\mathrm{s}}(t-t_0)\right)^{-\frac{3}{7}}
\end{equation}
and the density inside the shell is
\begin{equation}
\label{rho_t}
\rho_{\rm ion}(t)=\rho_0\left(1+\frac{7}{4}\frac{c_\mathrm{s,hot}}{R_\mathrm{s}}(t-t_0)\right)^{-\frac{2}{7}},
\end{equation}
where $\rho_0$ is the initial, constant density.
Here, $t_0$ is the time when the gas starts to react to the increase
in temperature and therefore pressure and starts expanding. It can be
approximated by the crossing time in the HII region, $t_0=R_{\rm
  S}/c_\mathrm{s,hot}$. 

We assume that stars form in the shell at a given time $t_{\rm
  SF}$. Thus, the stars inherit the position and velocity of the shock
front at that time. After formation they 
decouple from the gas and begin to move ballistically. Then, the stars
are only going to be  influenced by the gravitational potential 
which is given for the gas component by
\begin{equation}
V(r,t) = \frac{2}{3}\pi G\rho\left(r(t)^2-3r_{\rm max}^2\right)
\end{equation}
inside a spherically symmetric system. Here, $r_{\rm max}$ is the
maximal radius of the potential, which is the extend of the cloud and
$\rho$ is the mean density inside $r$.
Stars moving in the potential therefore are subject to an acceleration
\begin{equation}
a_{\rm spherical}(t)=-\nabla V(r) = -\frac{4}{3}\pi G \rho r(t).
\end{equation}
The most important point in our model is that for stars inside the
ionization and shock front $\rho = \rho_{\rm ion}(t)$, whereas for
stars outside of the front $\rho = \rho_0$. Thus, the stars inside the
HII-region feel less gravitational drag than the stars outside of it.
If there is additional mass close to the centre, i.e. a stellar
cluster of mass $M_{\rm central}$
associated with the ionization source, the resulting acceleration is
\begin{equation}
\label{tot_potential}
a_{\rm tot}(t)= a_{\rm spherical}(t) + a_{\rm central}(t) = -\frac{4}{3}\pi G
\rho r(t) - \frac{GM_{\rm central}}{r(t)^2}.
\end{equation}
\vspace{-.1cm}
\section{Trajectories}
\label{trajectories}
With these basic equations we can now calculate the trajectories of
the stars forming in the shell. From a given initial flux, $J_{\rm Ly}$, and number
density, $n_0$, we derive position and velocity of the front as well as
the density inside the HII-region according to Eqns
\ref{R_t}-\ref{rho_t}. For a star formed at a time $t_{\rm SF}$, this
provides the initial conditions
\begin{eqnarray}
r_0 &=& R_{\rm f}(t_{\rm SF})\\
v_0 &=& v_{\rm f}(t_{\rm SF}).
\end{eqnarray}
We then calculate the quantities at a given time $t_i=t_{\rm
  SF}+i\Delta t$ via
\begin{eqnarray}
\label{eq_motion}
r_{\rm i} &=& r_{\rm i-1} + v_{\rm i-1}\Delta t \\
a_{\rm i} &=& -\frac{4}{3}\pi G\rho r_{\rm i-1} - \frac{GM_{\rm central}}{r_{\rm i-1}^2} \\
v_{\rm i} &=& v_{\rm i-1} + a_{\rm i}\Delta t
\end{eqnarray}
(see Equation \ref{tot_potential}).

\begin{table}
\begin{center}
\begin{tabular}{llccc}
\hline
Case & Flux & $n_0$ & $T_{\rm ion}$ & $R_{\rm S}$ \\
& [$\flux$] & [$\ndens$] & [$K$] & [$\pc$]\\
\hline
1A & & $1\times 10^3$ & 6500& 0.21\\
1B & $10^{48}$ & $5\times 10^3$ & 7500 & 0.065\\
1C & & $1\times 10^4$ & 7750 & 0.038\\ \hline
2A & & $1\times 10^3$ & 7500 & 0.39 \\
2B & $10^{49}$ & $5\times 10^3$ & 8500 & 0.11 \\
2C & & $1\times 10^4$ & 8750 & 0.065 \\ \hline
3A & & $1\times 10^3$ & 8500 & 0.69 \\
3B & $10^{50}$ & $5\times 10^3$ & 9500 & 0.19 \\
3C & & $1\times 10^4$ & 9750 & 0.11 \\ \hline
\end{tabular}
\caption{Initial conditions for the different cases as calculated with Cloudy \citep{Ferland:1998fv}.\label{tab:ICs}}
\end{center}
\end{table}

We investigate three different fluxes, corresponding to a single
B-star, a single O-star and an O/B-association ($J_{\rm
  Ly}=10^{48}\flux$, $J_{\rm Ly}=10^{49}\flux$ and $J_{\rm
  Ly}=10^{50}\flux$). The number densities are $n_0=10^3\ndens$,
$n_0=5\times10^3\ndens$ and $n_0=10^4\ndens$, respectively. For
simplicity, the gas is assumed to be atomic ($\mu=1$). We use the
Cloudy code to estimate the initial $T_{\rm ion}$ and
$R_{\rm S}$. Calculations were performed with version 08.00 of Cloudy, last
described by \citet{Ferland:1998fv}. As the calculation of the
trajectories is very inexpensive, very small time steps can be used
for integration in order to minimize numerical errors due to the
discretization. A resolution study shows that $\Delta t = 1\yr$ leads to
converged orbital evolution that is independent of  $\Delta t $. 
The model parameters for the three cases are given in Table
\ref{tab:ICs}. A return of the proplyds or the stars is more likely the earlier they are formed
as they are still closer to the central potential. Therefore, we
investigate star formation times $t_{\rm SF}=200\kyr$ and 
$t_{\rm SF}=500\kyr$, as suggested from simulations \citep[e.g.][]{Gritschneder:2009lr}. 

The results are presented in Figure \ref{fig:100Msun} and Figure \ref{fig:1000Msun} for a
central mass of $M_{\rm central}=100\Msun$ and $M_{\rm
  central}=1000\Msun$, respectively. Let us first focus on Figure
1. Plotted is the position of the 
ionization front (dotted) as well as the trajectories for stars
forming after $200\kyr$ (solid) and $500\kyr$ (dashed). 
In all cases, the stars will overtake the ionized shell due to the
fact that the shell decelerates by sweeping up the surrounding
material while the stars feel just the gravitational force of the hot
interior. The stars, once outside of the shell experience a stronger
gravitational force by the high mass of the shell and and pulled
inwards again (e.g. the black line in 
the middle panel of Figure \ref{fig:100Msun}). In the cases with
higher density, the time spent outside the shell is very brief as the
total mass is higher and therefore the pull inwards is
stronger when the mass of the shell is added. These short
'excursions' beyond the shock front can lead to rather abrupt changes in
the trajectory (e.g. the blue line in the right panel of Figure
\ref{fig:100Msun}). 
As it can directly be seen in Figure \ref{fig:100Msun}, the
cases the stars do not return in time to the center in order to fit the age
of the proplyds. In fact they do not even return within the first
$\approx 5\Myr$. After  that time, the first supernovae will explode,
altering the system significantly\footnote{Note that \citet{Wright:2012kx} find
  proplyd-like objects at a distance of $6-14\pc$ of about 65 O-type
  stars in Cygnus. This could be explained by the high flux scenario
  (right panel of Figures \ref{fig:100Msun} and
  \ref{fig:1000Msun}). However, this still requires these objects to
  be older than $1-2\Myr$.}.
\begin{figure}
\begin{center}
{\centering 
\includegraphics[width=6cm]{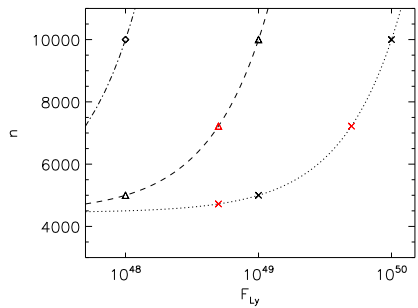}}
\end{center}
\caption{Flux versus density for three different groups of return
  times, adopting a central cluster mass of $1000\Msun$. Diamond/dash-dot:
  $0.5-1\Myr$, triangles/dashed: $1-2\Myr$ and crosses/dotted:
  $2-5\Myr$. Black symbols are the results from Figure
  \ref{fig:1000Msun}, red symbols are additional test to assess the
  predictive power of this figure.
    \label{fig:F_vs_n}}
\end{figure}

With an increased central mass (Figure \ref{fig:1000Msun}), the
situation changes. Focussing first on the left panel of Figure
\ref{fig:1000Msun}, one can see that stars triggered in such a dense
and massive stellar environment can indeed turn around within
$5\Myr$. 
However, this situation is quite unrealistic as the adapted flux
is only $10^{48}\flux$, whereas for central stellar masses of
$1000\Msun$ there should be at least one O-star, 
i.e. at least a flux of $10^{49}\flux$. This is the situation in the
middle panel of Figure \ref{fig:1000Msun}. Here, in the densest cases
the stars return within $1-2\Myr$, a 
long time before the first supernova explodes. However, those stars
would still be $0.8-1.5\Myr$ old, older than the proplyds are assumed
to be ($<0.5\Myr$, e.g. \citealt{Smith:2005fk}). As shown in the right panel of Figure
\ref{fig:1000Msun}, the situation becomes worse for even larger
fluxes. However, in a very dense environment (e.g. Orion), the return
time can be small enough (see below, \S \ref{Orion}).

In order to generalize our results, we divide the trajectories for the 
$M_{\rm central}=1000\Msun$ case into groups with a return time of
$0.5-1\Myr$, $1-2\Myr$ and $2-5\Myr$, respectively. We plot the flux
versus the density for these cases in Figure \ref{fig:F_vs_n} and fit
linearly for each age group. This figure directly allows to estimate
the return time for a given observed object. In these cases the fits
are $n = 4\times10^3[\ndens] + b\cdot F_{\rm Ly}[\fitnorm]$ with the
slope being  $b=5.5\times10^{-45}$, $b=5.5\times10^{-46}$ and
$b=5.5\times10^{-47}$, respectively. Remarkably, a change of return
time simply means changing the slope by an order of magnitude. In order
to test the predictive power, we perform additional calculations at
$F_{\rm Ly}= 5\times10^{48}\flux$ and $F_{\rm Ly}=
5\times10^{49}\flux$ at the respective densities given by the fit. The
resulting trajectories confirm a return in the expected time frame.

Altogether, our calculations show that triggered stars only return
within the first few $\Myr$ in the densest cases. Thus, it can be
readily explained that, if the proplyds have a triggered origin, they
only exist in a very dense subgroup of HII regions, e.g. in Orion (see
\S \ref{Orion}).
\vspace{-.1cm}
\section{Observability}
\label{currentstate}
\begin{figure*}
\begin{center}
{\centering 
\includegraphics[width=15cm]{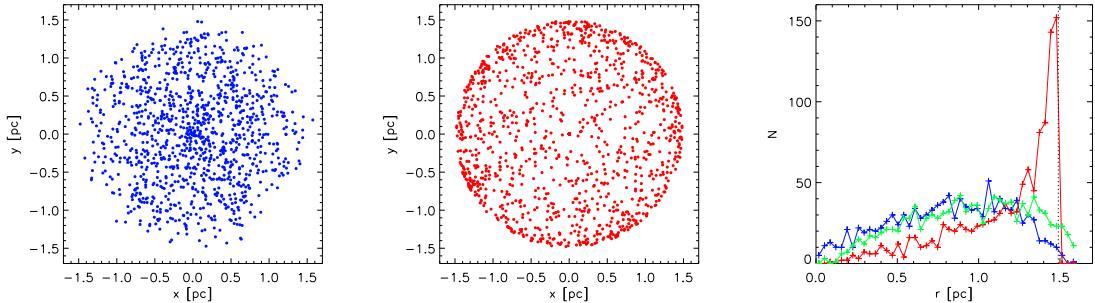}}
\end{center}
\caption{Left panel: Distribution of stars (random points) for a
  homogeneous sphere. Middle panel: distribution 
for a triggered shell. Right panel: comparison in
radial bins. Blue: in situ, red: triggered stars. In the right hand
panel a random distribution is over-plotted in green for
comparison. The realizations are done for 900 points (stars). The
triggered distribution has a distinct peak at the current position of
the shock front (dashed line).
    \label{fig:distribution}}
\end{figure*}
With the calculated trajectories, we can now investigate how an ensemble
of triggered stars differs from a randomly in situ formed generation
of stars. For reasons of simplicity, we assume stars are formed at a
constant rate from $200\kyr$ to $500\kyr$. We adopt 900 particles or
stars, about the order of magnitude of stars in the Central Orion
Nebula. For the in situ distribution, we adopt the most simple case that the stars are 
formed at a random location within $R=r_{\rm
  max}=1.5\pc$ and have a randomly orientated velocity of $1\kms$.
Then their position is evolved according to Eqns
\ref{eq_motion}. As the total stellar mass is only a small fraction of
the gas mass, we neglect their own self-gravity\footnote{Note that
  the reduced gravitational potential for particles inside the
  HII-region is still taken into account.}. 

For the triggered case, we assume the stars form at a random position
inside the current shock front and are subsequently subject to the
same potential as in the in situ case. 
In the following, we focus on Case
2B. Here, there is about $6000\Msun$ swept up in the shell and
$R=r_{\rm max}=1.5\pc$. Assuming a star formation efficiency of either
$3\%$ or $9\%$ and every star formed having either $0.2\Msun$ or
$0.6\Msun$, this would correspond to 900 
stars formed, in agreement with the number of points chosen. 
The results at $t=500\kyr$ are shown in Figure \ref{fig:distribution}
in a two-dimensional projection. The blue crosses
indicate the in situ case, and the red crosses the star formation
being triggered in a shell. 
In the right hand panel we show the distribution in radial
bins and we overplot the distribution for a random
distribution of stars in green in the right hand panel.

From Figure \ref{fig:distribution} the distinction is
very clear. The ring-like structure in the red crosses in panel
two is a clear feature of star formation, triggered in a shell-like
environment. Therefore it appears possible to distinguish between a
distribution of stars that was triggered compared to one that formed
randomly as long as enough stars are found. The ring is the result of
the relatively similar initial velocities of the stars formed in the
shell during the first $500\kyr$. 
In general, however, it is very likely that there is a mix of
both populations in all observed regions. 

In principle, the distribution peak of the triggered component makes
it possible to distinguish triggered stars from randomly formed stars.
Therefore, the two
components can be told apart, at least by number. We provide the coefficients for
a 4th order polynomial in Table \ref{tab:coeff}. 
\begin{table}
\begin{center}
\begin{tabular}{lcccccc}
\hline
Case & A0 & A1 & A2 & A3 & A4 \\ \hline
In situ & 0.041 & 0.29 & 0.56 & -0.87 & 0.031 \\
Triggered & 0.065 & -1.5 & 9.1 & -17.4 & 10.7 \\ \hline
\end{tabular}
\caption{Coefficients for 4th order polynomials fitted to the
  distributions in Fig \ref{fig:distribution}, right hand panel. The
  distributions were normalized to the maximum value of the triggered
  case for the fit. \label{tab:coeff}}
\end{center}
\end{table}
\vspace{-.1cm}
\section{Orion}
\label{Orion}
\begin{figure*}
\begin{center}
{\centering 
\includegraphics[width=5.5cm]{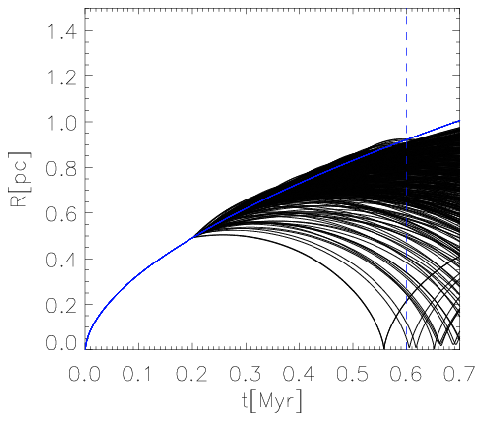}
\includegraphics[width=5.5cm]{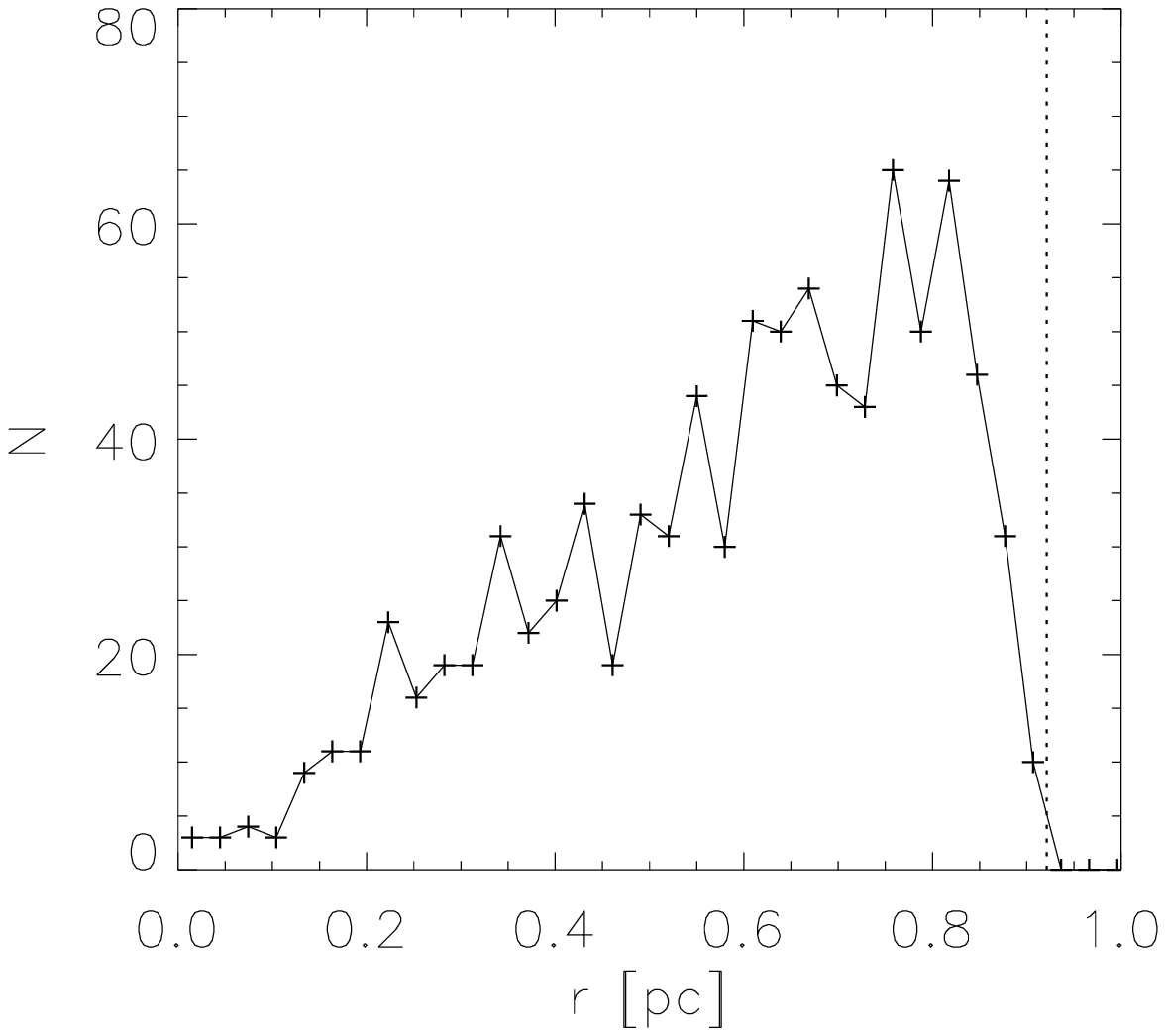}
\includegraphics[width=5.5cm]{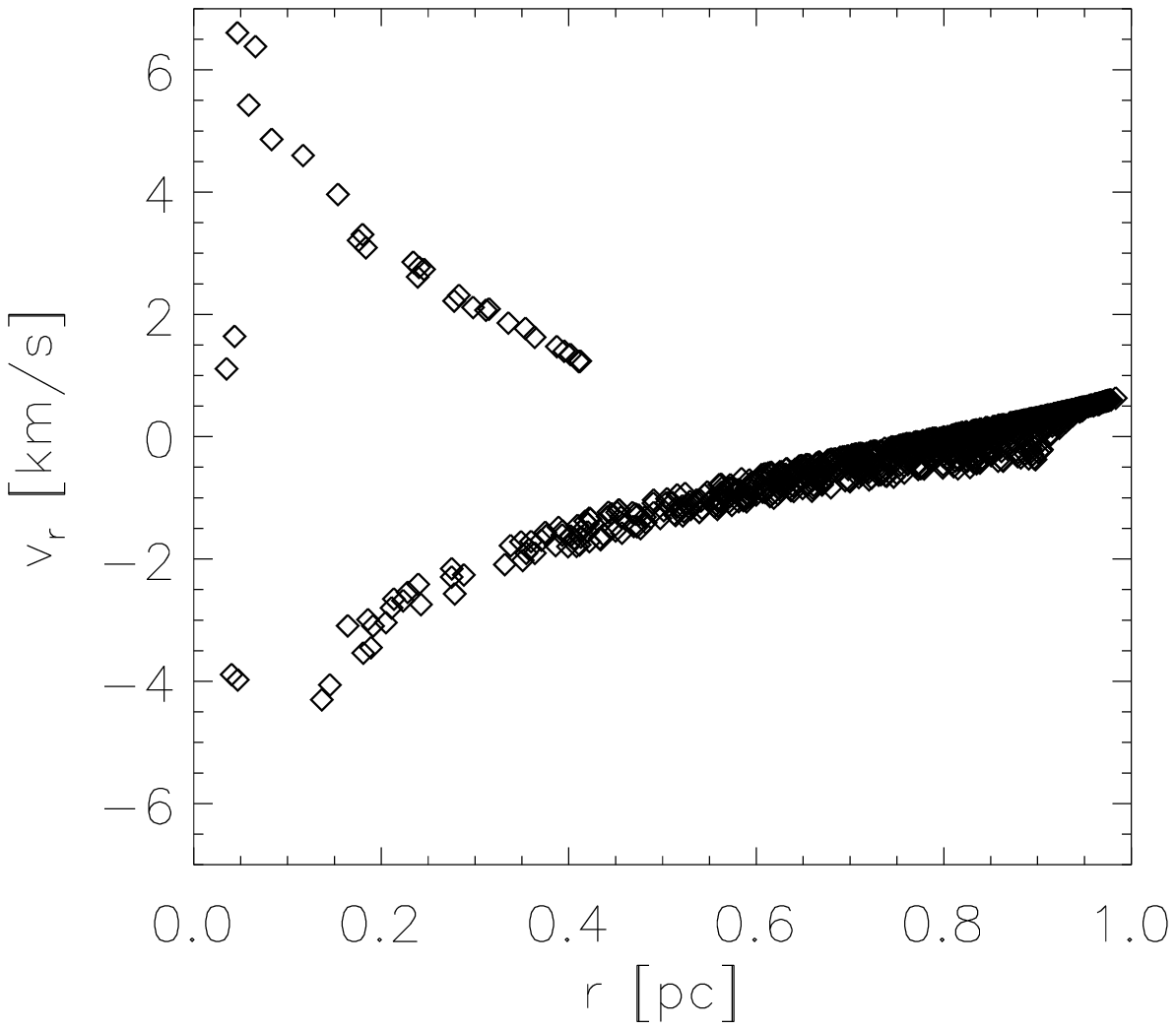}}
\end{center}
\caption{The special case of Orion: to explain the high current
  electron density we asume a high initial density, which in term
  leads to the return of triggered stars. Left panel: trajectories of
  the particles (stars). Blue solid line: position of the ionization
  front, blue dashed line: current age of $\theta$ Orions C. Middle
  panel: number of triggered stars in radial bins at t=
  $0.6\Myr$. Dashed line: current front 
  position. Right hand panel: radial velocity versus radius at t= $0.6\Myr$. A clear
  correlation with distance can be seen. The stars with positive velocities
  already had the closest encounter and are moving outwards again
    \label{fig:orion}}
\end{figure*}

As Orion is a very dense HII-region, it is worthwhile to investigate
that case in particular. The brightest star in the trapezium cluster,
$\theta$ Orionis C has a luminosity of $\approx10^{5.4}\Lsun$, a
surface temperature of $\approx45000\K$ and an age of less than $
0.6\Myr$ \citep {Howarth:1989,Donati:2002}. 
For our model, we adopt these values and assume the O-star has an age
of $0.6\Myr$, which corresponds to the age of this subgroup
\citep{Brown:1994}. To estimate the density at the onset of ionization
we use the current electron density in the outskirts of Orion,
where the gas is less perturbed by e.g. stellar winds. 
From \citet{Rubin:2011}, we take the current value to be about $350\ndens$.
Running a cloudy model with an initial density
of $n_0=4\times 10^4\ndens$ and the stellar values above yields a
similar density after $\approx 0.6\Myr$. The resulting initial
conditions are $R_{\rm S}=0.025\pc$ and $T_{\rm ion} = 10^4\K$.
 Note that these initial conditions are out of the range of Figure
 \ref{fig:F_vs_n} and in the region for short expected return times.
The total stellar mass in Orion can be estimated as $384\Msun$
\citep[e.g.][]{Hillenbrand:1997}. We calculate the trajectories for
these values and plot them in Figure \ref{fig:orion} (left hand
panel). It is immediately clear, that a large number of 
stars are on their way inwards or in close proximity to the trapezium
cluster, even after $0.6\Myr$ (blue dashed line). An interesting
feature is the reflection point at about $0.55\Myr$. Here, the first
returning stars have their closest encounter with the central
potential. From then on the population is mixed and there are proplyds
moving inwards and outwards. In the middle panel
we show the distribution of the stars at $t=0.6\Myr$. Although the
peak of the distribution is still at the position of the ionization
front, there is a significant number of triggered stars inside of this
radius. Therefore, the assumption of a triggered origin for the
proplyds is definitely valid. The distribution shown here is the distribution of all
  triggered stars. For the proplyds, the distribution should be
  strongly cored and drop with radius as $R^{-2.58}$, as shown
  by \citet{Henney:1998}. As we are not investigating the interaction
  of the stellar feedback with the triggered 
  stars, our model can not predict the distribution of the proplyds.
A final test for this scenario are the
proper motions of the triggered stars. In Figure \ref{fig:orion}
(right hand panel), we plot the radial velocity versus the
radius. As expected, there is a strong correlation with distance. The further out
the stars are, the slower they are, whereas closer to the centre they
move much faster, as their kinetic energy is close to maximal. 
In addition, some stars have positive radial velocity, they already
had their closest encounter and are moving outwards. Thus,
our conjecture could be conclusively proven by a detailled analysis of the
proper motion of proplyds and triggered stars.

The parameters chosen for Orion are the combination of an older star
and a lower density as inferred from from the outskirts of Orion. In
general, several combinations of density and stellar age are
possible, e.g. a younger star and a correspondingly denser
medium, which would result in an earlier return time (cf \S \ref{trajectories}).
The return time and the assumed age of $\theta$ Orions C of $0.6\Myr$
is slightly larger than the estimated lifetime of the proplyds of 
about $0.5\Myr$. However, if the stars return after spending most of
their lifetime further away from the ionizing source (Figure
\ref{fig:orion}), their life-time will be longer than the ones
estimated from their current, higher photo-evaporation in close
proximity. 
The age determination is a main uncertainty, e.g. \cite{Jeffries:2011}
adopt a higher age for the subgroup, but do not exclude a smaller age
like the one adopted here \citep[as suggested by
e.g.][]{Smith:2005fk,Tobin:2009}. For our model, the challenge is to explain the
fast return, a slower return, i.e. a higher age is not a problem.
We therefore conclude, that the proplyds in Orion
are indeed the smoking gun of a population of stars triggered in an
ionization shell. 
\vspace{-.5cm}
\section{Discussion \& Conclusions}
\label{conclusions}
In general, the stars inside HII-regions are
most likely not stars triggered within the shell of the ionization
front, as they would have formed with too high radial velocities to
have returned yet. 
However, in dense cases like Orion, the HII region confinement is
strong enough to explain the observed proplyds by our model of
ionization triggered star formation with a subsequent return.

The HII region expands rapidly and the in situ and triggered
populations of stars will be mixed, especially after the reflection
point (see \S \ref{Orion}), when triggered stars are moving inwards
and outwards. Therefore, we do not expect an 
age spread proceeding with the speed of the ionization front. Instead,
we propose a sequence which shows a radial dependence, and which
results from different levels of harassment. Closest in are the proplyds, where the
disk of a young star interacts directly with the feedback of the
massive star. In the medium range, there are Class II protostars,
where the ionization already was able to remove the envelope
partly. At the outskirts of the HII region, there are Class I stars, which either have
been triggered by the ionizing radiation or pre-existed and have not
lost their envelope yet, as the feedback is much weaker further away.

Whether the protostars together with the central mass form a
bound or an unbound cluster depends on the precise initial
conditions and has to be determined with e.g. N-Body simulations. 
The triggered stars here, however, are likely to be loosely bound or
will be removed due to their high initial velocity. In
addition, the spherically symmetrical potential gets diminished by the
gas loss, so the tidal radius shrinks. Furthermore, as soon as the
central stars explode in a supernova, additional gas is going to be
removed, leading to an even shallower potential. 

We have shown that a triggered population shows a different density
distribution compared e.g. to a randomly formed distribution of
stars. 
Especially, the triggered distribution shows a characteristic peak at the
current position of the shock front. Inside the HII-region, the
stellar component is dominated by in situ stars.
For the densest regions like Orion, enough triggered stars return
within $0.6\Myr$ to provide a feasible explanation for the origin of
the proplyds as well as their exclusiveness.
\vspace{-.6cm}
\section{Acknowledgements}
We thank the referee for valuable comments on the manuscript.
M.G. acknowledges funding by the Alexander von Humboldt Foundation in
form of a Feodor-Lynen Fellowship. Part of A.B.'s research was supported by
the Excellence Cluster "Origin and Structure of the Universe".
\vspace{-.6cm}
\bibliographystyle{mn2e}
\bibliography{references}
 
\end{document}